\documentclass[journal=jctcce,manuscript=article,layout=onecolumn]{achemso}
\setkeys{acs}{articletitle=true,doi=true,etalmode=truncate,maxauthors=10}
\usepackage[version=4]{mhchem} % Formula subscripts using \ce{}
\usepackage{textcomp}
\usepackage{graphicx}
\usepackage{color}
\usepackage{soul}
\usepackage{dcolumn}
\usepackage{bm}
\usepackage{amsmath,amssymb}
\usepackage{natbib}
\usepackage{array}
\usepackage{supertabular}
\usepackage{hhline}
\usepackage{hyperref}
\usepackage{lipsum}
\newcommand{\doi}[1]{\href{http://dx.doi.org/#1}{\nolinkurl{#1}}}

\newcommand{\RED}[1]{#1}

\title{Melting behavior of \ce{CaO} at high temperature and pressure: a molecular dynamics study}
\author{Francesca Menescardi}
\affiliation{Dipartimento di Scienze della Terra, dell'Ambiente e della Vita (DISTAV), University of Genova, corso Europa 26, 16132 Genova, Italy}
\alsoaffiliation{Present address: Scuola Internazionale Superiore di Studi Avanzati (SISSA), via Bonomea 265, 34136 Trieste, Italy}

\author{Davide Ceresoli}
\email{davide.ceresoli@cnr.it}
\affiliation{Consiglio Nazionale delle Ricerche, Istituto di Scienze e Tecnologie Chimiche ``G. Natta'' (CNR-SCITEC), via Golgi 19, 20133 Milan, Italy}

\author{Donato Belmonte}
\email{donato.belmonte@unige.it}
\affiliation{Dipartimento di Scienze della Terra, dell'Ambiente e della Vita (DISTAV), University of Genova, corso Europa 26, 16132 Genova, Italy}

\begin{document}
%%%%%%%%%%%%%%%%%%%%%%%%%%%%%%%%%%%%%%%%%%%%%%%%%%%%%%%%%%%%%%%%%%%%%%%%%%%%%%%%%%%%%%%%%%%%%%%%%%
\begin{tocentry}
  \centering
  \includegraphics[width=8.4cm]{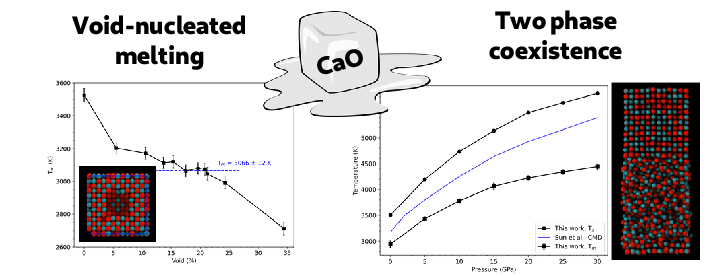}
%The surrounding frame is 9\,cm by 3.5\,cm, which is the maximum
%permitted for  \emph{Journal of the American Chemical Society}
%graphical table of content entries. The box will not resize if the
%content is too big: instead it will overflow the edge of the box.
\end{tocentry}

%%%%%%%%%%%%%%%%%%%%%%%%%%%%%%%%%%%%%%%%%%%%%%%%%%%%%%%%%%%%%%%%%%%%%%%%%%%%%%%%%%%%%%%%%%%%%%%%%%
\begin{abstract}
%%%%%%%%%%%%%%%%%%%%%%%%%%%%%%%%%%%%%%%%%%%%%%%%%%%%%%%%%%%%%%%%%%%%%%%%%%%%%%%%%%%%%%%%%%%%%%%%%%
The thermodynamic behavior of calcium oxide (\ce{CaO}) under high temperature and pressure conditions is critical for understanding the physics of planetary interiors. This study employs molecular dynamics (MD) simulations, including both classical and ab-initio approaches, to investigate the melting behavior of CaO. We calculate the melting temperature of \ce{CaO} by the void-nucleated melting and two-phase coexistence techniques, aiming to resolve discrepancies in experimental data on the melting point, which range from 2843~K to 3223~K in different studies due to the high reactivity and vapor pressure of the substance. The obtained results are $T_f = 3066\pm12$~K and $T_f = 2940\pm65$~K using the void-nucleated melting and the two-phase coexistence method, respectively. Additionally, we calculate the enthalpy of fusion and the high-pressure melting curve, for the first time without making any assumption on the Clapeyron slope. This is extremely important since in experiments the Claperyon slope of the melting curve is estimated from low pressure measurements and the overheating ratio (i.e. $\eta=\frac{T_s}{T_f}-1$, where $T_s$ represents the thermal instability limit corresponding to the homogeneous melting temperature of the solid) is often assumed to be constant in simulations. Our MD results show that $T_s$ increases more rapidly with pressure than $T_f$ and thus that the overheating ratio sensibly depends upon pressure. These findings contribute to the accurate modeling of the CaO phase diagram, which is essential for geochemistry, cosmochemistry, and materials science.
%%%%%%%%%%%%%%%%%%%%%%%%%%%%%%%%%%%%%%%%%%%%%%%%%%%%%%%%%%%%%%%%%%%%%%%%%%%%%%%%%%%%%%%%%%%%%%%%%%
\end{abstract}
%%%%%%%%%%%%%%%%%%%%%%%%%%%%%%%%%%%%%%%%%%%%%%%%%%%%%%%%%%%%%%%%%%%%%%%%%%%%%%%%%%%%%%%%%%%%%%%%%%

%%%%%%%%%%%%%%%%%%%%%%%%%%%%%%%%%%%%%%%%%%%%%%%%%%%%%%%%%%%%%%%%%%%%%%%%%%%%%%%%%%%%%%%%%%%%%%%%%%
\section{Introduction}\label{sec:introduction}
%%%%%%%%%%%%%%%%%%%%%%%%%%%%%%%%%%%%%%%%%%%%%%%%%%%%%%%%%%%%%%%%%%%%%%%%%%%%%%%%%%%%%%%%%%%%%%%%%%
Understanding the high-temperature and high-pressure thermodynamic behavior of \ce{CaO} is of paramount importance for the physics of planetary interiors since \ce{CaO} is one of the main constituent oxides of the silicate mantle of rocky planets like Mars, Mercury and the Earth~\cite{Anderson1989}. In particular, modeling the phase diagram of Ca-bearing silicate systems requires an accurate knowledge of the thermodynamic properties of the end-members, \ce{CaO} and \ce{SiO2}, such as their melting curve $T_f(P)$ and enthalpy of fusion.~\cite{Belmonte2017a} Outside the field of Earth sciences, \ce{CaO} (or lime, its name as a mineral) has well known technological applications in refractory ceramics and building materials such as cement.~\cite{Taylor1997}

Despite the popularity of this material both for research and practical purposes, the melting point of calcium oxide is still widely debated in the scientific community. The very high melting point ($>$2800 K) and the very high vapor pressure~\cite{Manara2014} make it hard to perform experiments on the liquid phase of \ce{CaO}. 

The experimental melting point ($T_f$) of \ce{CaO}, in fact, ranges from 2843~K to about 3223~K according to different measurements (see Ref.~\cite{Liang2018} and references therein). The high reactivity of \ce{CaO} with respect to water is believed to be one of the main reasons why the melting point of this material is so sensitive to the experimental conditions and setup, which makes values of $T_f$ that can be found in literature highly uncertain. Put in contact with \ce{H2O}, lime easily transforms into calcium hydroxide, i.e. \ce{Ca(OH)2}, introducing more difficulties in properly setting up the experiments and making \ce{CaO} the hardest refractory oxide to study. Similarly the reaction with \ce{CO2} leads to the formation of \ce{CaCO3}. In addition to that, pyrometric measurements (which are strongly affected by the \ce{CaO} vapor pressure) could lead to the underestimation of the melting temperature due to likely eutectic reaction with the tungsten crucible~\cite{Foex1965,Yamada1986}. Finally, performing experiments in oxidizing or reducing environments could determine variations in the observed melting point up to 30-50~K~\cite{Manara2014}.

Keeping these experimental issues in mind, the latest works on this topic by Manara et al.~\cite{Manara2014} and Bgasheva et al.~\cite{Bgasheva2021} employed laser heating techniques in different environments to shed new light on the melting point of lime, obtaining a value of $T_f = 3222\pm25$~K and $T_f = 3160\pm10$~K respectively, which, despite being about 60~K apart, reliably set the melting temperature of \ce{CaO} above the threshold of 3000~K.

Given the experimental difficulties at ambient pressure, to the best of our knowledge, the high-pressure melting curve of \ce{CaO} has not been measured yet. Therefore one of the most practical and efficient ways to extract information about the HT-HP thermodynamic properties of solid and liquid \ce{CaO} are atomistic simulations, in particular Molecular Dynamics (MD).

A first attempt to fill this gap was made by Seo et al.~\cite{Seo2004} which approached the \ce{CaO}-\ce{SiO2} system from a computational point of view in order to build a reliable phase diagram starting from the thermodynamic and structural properties of both the liquid and solid phases. Employing classical MD simulations, they obtained a melting temperature for \ce{CaO} of $3210\pm10$~K and an enthalpy of fusion ($\Delta H_f$) of $74.5$~kJ/mol. The value of $T_f$, though, was deduced from the \RED{thermal stability limit $T_s$} of the perfect bulk crystal, which is known to overestimate the actual melting point by several hundreds of K -- usually about 20\%~\cite{Belonoshko2006}. \RED{The thermal stability limit $T_s$ can be thought as the temperature at which the solid can't resist without breaking, i.e. when the free energy of formation of glide dislocations becomes negative so that they fill the entire solid.~\cite{Wilsdorf1965}} Thus, these results can be considered as inaccurate and will be disregarded for the purposes of comparison. A different approach was used by Alvares et al.~\cite{Alvares2020}, who employed classical MD to calculate the melting point of \ce{CaO} using the void-melting technique and ab-initio MD to simulate the thermodynamic and structural properties of the system. Their study resulted in a melting point of $T_f = 3156\pm10$~K and enthalpy of fusion of $\Delta H_f = 80.89$~kJ/mol, which are in better agreement with experimental data and can be considered as the computational reference for this work. However, the void-melting technique, which consists in creating an empty cavity in the bulk solid, is not justified by sound physical arguments. In fact, the nucleating defects in solids might be quite different from spherical voids carved into the bulk.

More recently Wang et al.~\cite{Wang2023}, using classical MD with a different interatomic potential, obtained a melting temperature of 2767~K, on the low end of the reported experimental range. Lastly, it is worth mentioning that an attempt to build a full phase diagram of the \ce{MgO}-\ce{CaO} binary system employing neural network interatomic potentials~\cite{Lee2022} led to a predicted melting temperature for the \ce{CaO} end-member phase of $T_f = 3057$~K.

Our work aims to take a step forward in the investigation of the \ce{CaO} system. First, we compare the void-melting and the two-phase coexistence methods at ambient pressure. Since the void-melting technique cannot withstand very high pressures, we then employ the two-phase coexistence method to calculate the melting curve of \ce{CaO} up to 30~GPa.

The only previous attempt to calculate the melting curve of \ce{CaO} was made by Sun et al.~\cite{Sun2010}. In that work, the authors calculated by classical MD the thermal instability curve of \ce{CaO} (i.e. the homogeneous melting temperature, $T_s$) up to 60 GPa and then inferred the melting curve scaling $T_s$ by a constant factor of 0.7, independently of pressure conditions. Thus, the high-pressure melting curve is deduced first by making an empirical assumption that fits $T_s$ to the assessed value of $T_f = 3200\pm50$~K at ambient pressure according to the NIST-JANAF Thermochemical Tables~\cite{Chase1998} and then applying the inferred scaling factor to all the computed values for $T_s$ on the thermal instability curve regardless of the pressure \RED{conditions}. As we will show in the Results section, the $\eta=\frac{T_s}{T_f}-1$ overheating ratio is not constant as a function of pressure and temperature and therefore the empirical results obtained on the melting curve of \ce{CaO} by Sun et al.~\cite{Sun2010} seem to be affected by large uncertainties.

In this work, we employ both classical and GPU-accelerated ab-initio MD. Since the computational cost of ab-initio MD is several orders of magnitude higher than classical MD, we employ ab-initio MD to extract the thermodynamic properties of single phases (i.e. solid and liquid \ce{CaO}), whereas we employ classical MD to simulate two-phase systems with interfaces (i.e. the solid-vacuum interface in the void-melting and the solid-liquid interface in the coexistence method). For classical MD simulations we use the same interatomic potential used by Alvares et al.~\cite{Alvares2020}. The reason is twofold. First, as shown in their paper, this interatomic potential reproduces very well the structural and dynamic properties of liquid \ce{CaO} calculated with ab-initio MD using the meta-GGA SCAN functional~\cite{Sun2015}. Second, by making a direct comparison with the results of Alvares et al.~\cite{Alvares2020}, we show how the technique and the details of the MD simulation impact on the calculated melting temperature. The final goal is to provide new physical insights on thermodynamic properties of the \ce{CaO} end-member liquid phase, which is a major component of chemical systems relevant to metallurgy, ceramics, geochemistry and cosmochemistry, to better constrain multi-component phase diagrams in a broad range of P-T conditions. In fact, several attempts was made in the past to assess melting phase relations in \ce{CaO}-rich silicate systems by computational thermodynamics~\cite{Taylor1990,Hillert1990,Eriksson1993} but only few of them make use of first principles DFT/MD results~\cite{Belmonte2017a} or extend their validity to high pressure conditions~\cite{Hudon2005,Belmonte2017a}. Ab initio-assisted computational thermodynamics is thus proposed as a modern approach to theoretical prediction of complex phase diagrams~\cite{Lee2022,Chew2023}.

In addition to the void-nucleated and the two-phase coexistence, there exist other MD methods to simulate melting: they include \RED{the extrapolation of cluster melting~\cite{Pahl2008}}, the Z-method~\cite{Belonoshko2006,Alfe2011} with its modification~\cite{Wang2013}, the interface pinning method~\cite{Pedersen2013}, and the direct calculation of solid and liquid free energy using thermodynamic integration~\cite{Kirkwood1935,FrenkelSmit}. All these methods display limitations which have been extensively discussed in the literature~\cite{Alavi2005,Zhang2012}. For instance, \RED{the extrapolation of the melting point of finite clusters tends to the surface melting temperature, which in the vast majority of solids is lower that the bulk $T_f$.} The Z-method suffers from hysteresis and might lead to erroneous predictions on the melting curve if a solid-solid phase transition occurs at HT-HP conditions~\cite{Belonoshko2013}. Free energy methods, which requires highly accurate calculation of the liquid phase, are rigorous in principle, but their application to complex multi-component systems is still challenging and expensive, hence not straightforward~\cite{Hong2022}. The direct methods employed in this work are easier to implement and have been successfully applied to determine the melting behavior of a wide range of solids. Moreover, the combination of classical and ab-initio MD calculations has already proved to be a promising computational strategy in predicting the melting curve of multi-component systems up to extreme P-T conditions~\cite{DiPaola2016}.

%%%%%%%%%%%%%%%%%%%%%%%%%%%%%%%%%%%%%%%%%%%%%%%%%%%%%%%%%%%%%%%%%%%%%%%%%%%%%%%%%%%%%%%%%%%%%%%%%%
\section{Computational methods}
%%%%%%%%%%%%%%%%%%%%%%%%%%%%%%%%%%%%%%%%%%%%%%%%%%%%%%%%%%%%%%%%%%%%%%%%%%%%%%%%%%%%%%%%%%%%%%%%%%
\subsection{Classical Molecular Dynamics}\label{sec:CMD}
We perform classical MD (CMD) simulations using the Born-Meyer-Huggins potential~\cite{Huggins1933} to describe the interaction between atoms. The BMH empirical potential is expressed as a function of the distance between two atoms and is written as follows:
\begin{equation}
  V(r_{ij}) = \frac{1}{4\pi\epsilon_0}\frac{q_i q_j}{r_{ij}} + 
              A_{ij}\exp\left(\frac{\sigma_{ij}-r_{ij}}{\rho_{ij}}\right) -
              \frac{C_{ij}}{r_{ij}^6} + \frac{D_{ij}}{r_{ij}^8} \label{eq:BHM}
\end{equation}
\RED{In this equation, $i$ and $j$ correspond to atoms pairs, while $r_{ij}$ represents the interatomic distance between the two atoms. The first term is the Coulomb interaction, $q_i, q_j$ are the ionic charges and $\epsilon_0$ is the vacuum dielectric constant. The second term describes the short range Pauli repulsion. The third and fourth term describe the long-range van der Waals and London dispersion terms.} The parameters of the interatomic potential are listed in tab.~\ref{tab:parameters} and are the same parameters used in previous works~\cite{Bouhadja2013,Alvares2020} to successfully describe Ca-O atomic interactions.

\begin{table*}\begin{center}
\begin{tabular}{cccccc}
\hline\hline
Atom pair & $A_{ij}$ (kcal/mol) & $\rho_{ij}$ (Å) & $\sigma_{ij}$ (Å) & $C_{ij}$ (kcal/mol Å$^6$) & $D_{ij}$ (kcal/mol Å$^8$)\\
\hline
Ca--Ca & 0.080600 & 0.0800 & 2.3440 & 483.27068 & 0 \\
Ca--O  & 0.177320 & 0.1780 & 2.9935 & 973.0907  & 0 \\
O--O   & 0.276344 & 0.2630 & 3.6430 & 1959.372  & 0 \\
\hline\hline
\end{tabular}
\caption{Parameters for the Born-Meyer-Huggins empirical potential of \ce{CaO}. The ionic charge of Ca is $+$1.2\,e,while the charge of O is $-$1.2\,e.}\label{tab:parameters}
\end{center}\end{table*}

Note that the BMH potential is attractive at very short distances for the cation-anion pairs. This attractive well is unphysical and whenever a Ca-O pair gets too close, the molecular dynamics run is spoiled. This issue becomes very serious at high temperatures (i.e. when the atoms move very fast) and at high pressure (i.e. when the atoms are forced closer to each other). One solution is to reduce the timestep in order to ensure that the equations of motion are integrated accurately. Unfortunately this cannot avoid that there exists an upper bound on the pressure that can be applied. Another accepted solution is to add a steep repulsive 24-6 Lennard Jones potential to insure the Pauli repulsion at short distances. We decided to use a timestep smaller than that used in Ref.~\cite{Alvares2020} and to limit the range of pressure, without adding any steep repulsive potential.

We employ the LAMMPS code~\cite{Plimpton1995,Thompson2022} to perform classical MD. Long-range Coulomb interactions are treated within the Ewald summation method and periodic boundary conditions are applied in all three directions. Supercell dimensions and total simulation time vary depending on the different techniques that we use (i.e. the void-nucleated melting or the two-phase coexistence method, see next Section), and all simulations are performed in the isobaric (NPT) ensemble, employing the Nosé-Hoover barostat and thermostat~\cite{Hoover1985}, except otherwise stated. We always use a time step of 0.5~fs for a minimum of 200,000 steps which makes the minimum total simulation time of 100 ps. With this empirical potential, we are able to run simulations of much larger supercells, which, depending on the technique employed, contain a minimum of 1728 atoms and a maximum of 5832 atoms.

\subsection{Ab-initio Molecular Dynamics}\label{sec:AIMD}
We carried out Density Functional Theory (DFT) calculations to carry out Born-Oppenheimer ab-initio molecular dynamics (AIMD) simulations both on the solid and the liquid phases, \RED{to calculate their enthalpy as a function of temperature.} We used the GPU-enabled Quantum Espresso v6.8~\cite{Giannozzi2009,Giannozzi2017,Giannozzi2020}. We employ standard norm-conserving pseudopotentials~\cite{Troullier1991} with a wavefunction cutoff of 70 Ry. We employ PBEsol~\cite{Perdew2008} exchange-correlation functional, which predicts fairly well the equations of state of a large number of inorganic materials. In fact, PBEsol lies in between LDA which overbinds solids (i.e. smaller volume, larger bulk modulus) and PBE which slightly underbinds solids (i.e. larger volume, smaller bulk modulus). The PBEsol functional has been shown to predict thermodynamics properties such as enthalpies of formation in good agreement with experiments~\cite{Csonka2009}.

%%%%%%%%%%%%%%%%%%%%%%%%%%%%%%%%%%%%%%%%%%%%%%%%%%%%%%%%%%%%%%%%%%%%%%%%%%%%%%%%%%%%%%%%%%%%%%%%%%
\section{Results and Discussion}\label{sec:results}
%%%%%%%%%%%%%%%%%%%%%%%%%%%%%%%%%%%%%%%%%%%%%%%%%%%%%%%%%%%%%%%%%%%%%%%%%%%%%%%%%%%%%%%%%%%%%%%%%%
\subsection{Melting temperature of CaO}
From the thermodynamic point of view, the melting point $T_f$ of a crystal is defined as the temperature at which the molar Gibbs free energies of the solid phase and the liquid phase are equal at any given pressure. Heating of a crystal beyond $T_f$ without any onset of melting is called overheating, which has an upper limit above which the lattice structure collapses. This limit defines the mechanical melting point, or thermal instability temperature $T_s$ and it is very hard to reach experimentally. The reason is that in real crystals the abundance of lattice defects, free surfaces and grain boundaries, which are statistically very common in the majority of the samples, help nucleate the melting of the crystal very close to $T_f$, which is usually about 20-25\% lower than $T_s$~\cite{Jin2001}. An alternative definition of $T_f$ is the temperature at which a solid and liquid phase of the same composition coexist at equilibrium (congruent melting). These two observations suggest two strategies to determine the melting temperature of a solid: the \textit{void-nucleated melting} and the \textit{two-phase coexistence} technique.

\subsubsection{Void-nucleated melting technique}
The basic idea behind this technique is that simulations performed on perfect crystal structures to simulate homogeneous melting inevitably produce overheating, due to the complete lack of extrinsic defects. Thus, simulations performed on ideally perfect crystals are useful only in the attempt to find $T_s$, but never give a reliable $T_f$ unless a defect is deliberately introduced in the input crystal structure. The first attempt to investigate the role of lattice defects in determining an accurate melting temperature for crystals was performed by Lutsko et al.~\cite{Lutsko1989} and since then this concept was successfully applied in many other works~\cite{Solca1997,Solca1998,Agrawal2003,Jakse2005}. With this technique, a ``void'' is created in the center of the relaxed input crystal by removing an atom and a certain number $n$ of its neighbors so that the melting process will be nucleated. The system is then gradually heated up at ambient pressure and at a certain heating rate (typically $\sim 10^{12}$~K/s for this kind of MD simulation) until it melts completely. The melting is observed as a sharp discontinuity in both the volume and the enthalpy values of the system.

This procedure is repeated increasing the defect dimensions simulation after simulation, going from $n=0$ (which represents the perfect crystal and will output the thermal instability temperature $T_s$) up to a void dimension which for \ce{CaO} corresponds to 35\% in volume with respect to the supercell. After this limit, in fact, we observe the collapse of the crystal structure at the very beginning of the simulation, meaning that the defect is too big for the lattice to be stable. The resulting $T_f$ is expected to gradually lower as the volume of the defect increases up until a \textit{plateau} is reached, which means that the melting temperature of the crystal is independent from the defect dimensions. That plateau temperature thus represents the desired reliable melting temperature of the crystal, while the values found before and after the plateau can be referred to as \textit{apparent} melting temperatures.

\begin{figure}\begin{center}
  \includegraphics[width=0.9\columnwidth]{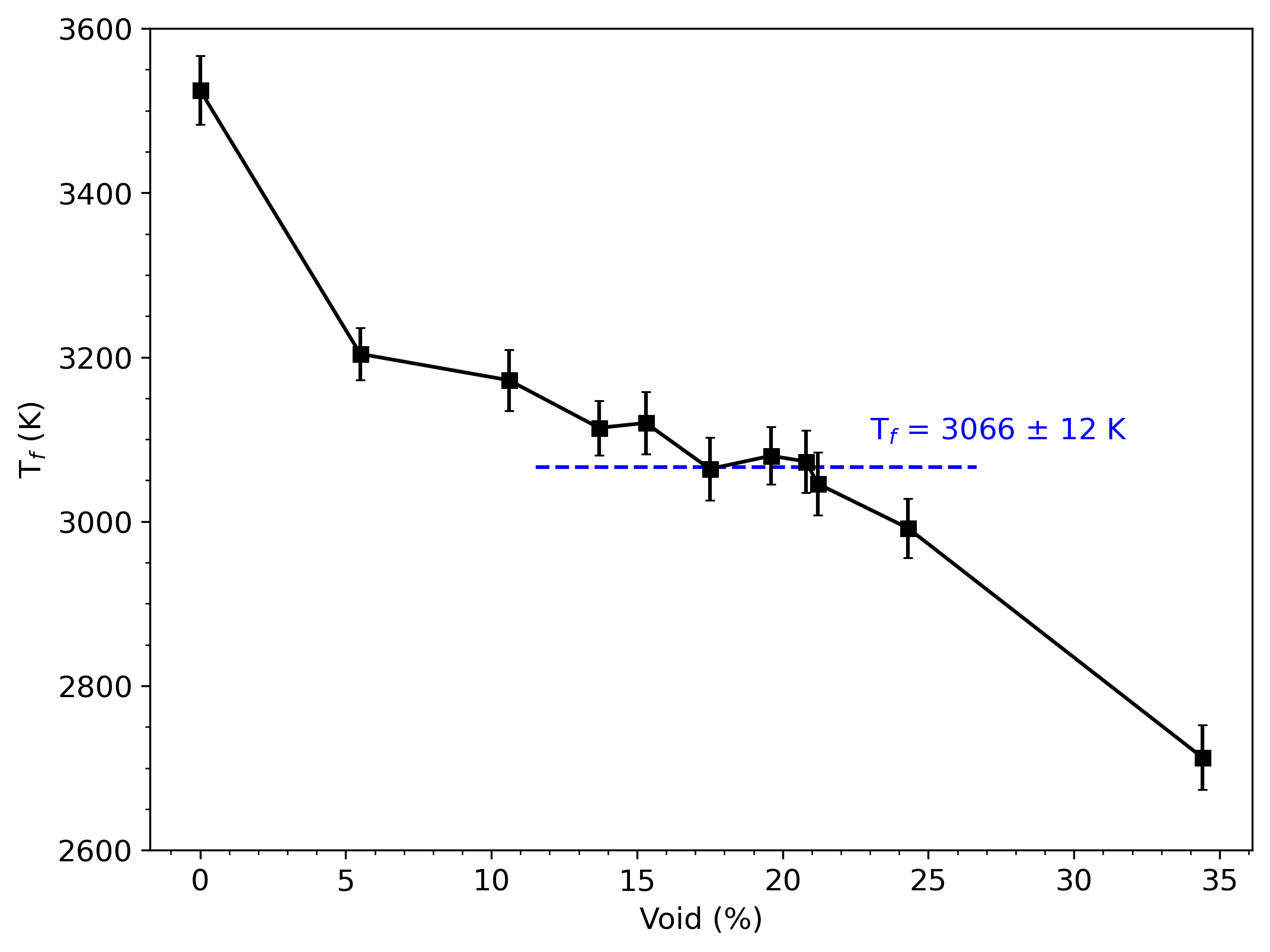}
  \caption{Apparent melting temperature of crystal \ce{CaO} as a function of the defect volume (in \% of the simulation box volume). The plateau melting temperature is reached at $3066\pm12$~K, which represents the true melting temperature $T_f$ of the solid.}\label{fig:void}
\end{center}\end{figure}

We run several NPT simulations at ambient pressure on a 9$\times$9$\times$9 supercell of \ce{CaO} that contains 5832 atoms, gradually increasing the dimension of a single spherical void placed in the center of the simulation box. Fig.~\ref{fig:void} shows the decrease in the apparent melting temperature of the crystal as a function of the void size in volume percent. In the range of defect size between 17\% and 22\%, though, the values of $T_f$ reach a plateau which corresponds to a melting temperature of $3066 \pm 12$~K. Increasing the defect dimensions above 22\% in volume causes instability in the crystal structure, resulting in $T_f$ values that are too low to be realistic.

Given that there is a wide range of uncertainty on the experimental melting temperature of solid \ce{CaO}, the value we obtained falls not so far from the most recent experimental values of $T_f=3222\pm25$~K and $T_f=3160\pm10$~K~\cite{Manara2014,Bgasheva2021}. On the other hand, compared to previous computational works, our melting temperature is found to be slightly lower than the value obtained by Alvares et al. ($3156\pm10$~K) employing the very same technique~\cite{Alvares2020}, but it is perfectly aligned with the melting temperature obtained by Lee et al. ($T_f=3057$~K) employing neural network interatomic potentials~\cite{Lee2022}.

Note that using the same void-nucleated technique but with a different interatomic potential fitted to first principles LDA and GGA calculations, Wang et al.~\cite{Wang2023} obtained a much lower melting temperature of 2775~K, which proves the effect an inaccurate interatomic potential could have on melting point simulation of refractory materials like \ce{CaO}. Therefore our result certifies a melting temperature for \ce{CaO} above 3000~K, confirming that any value of $T_f$ previously reported at temperatures around 2800~K should be disregarded.

\subsubsection{Two-phase coexistence technique}
The two-phase solid-liquid coexistence technique was first conceived as a computational strategy to determine the melting curve of aluminum employing molecular dynamics~\cite{Morris1994}. Since then, this technique was successfully applied in more recent times by Agrawal et al.~\cite{Agrawal2003} and was then further developed by Hong et al.~\cite{Hong2013}.

With this technique the simulation box is constructed by doubling the supercell along one axis. The first half of the box consists of a solid crystal phase, while the other half is filled with its corresponding liquid phase. Our MD simulation is performed on a 6$\times$6$\times$12 supercell of 3456 atoms, half of which are in the solid phase and the other half in the liquid phase. In practice this is done by imposing two different temperatures on the two halves. One half of the simulation is kept close to the estimated $T_f$, while the other half is heated to very high temperature until it melts. Then its temperature is gradually reduced towards the estimated $T_f$. To apply pressure we fix the lattice parameter parallel to the solid-liquid interface according to its average value at the desired P and T conditions, and we apply the Berensden barostat~\cite{Berendsen1984} along the perpendicular direction. This is because the in-plane components of the stress tensor include an extra term due to the interfacial tension.

First, we run a series of NPT short simulations (i.e. 10~ps) to equilibrate the system at any desired conditions of temperature and pressure, to determine the lower and upper limits for $T_f$. If the system crystallizes then $T<T_f$. If the system melts completely, clearly $T>T_f$. As the temperature approaches $T_f$ the solid-liquid interface moves increasingly slower and it is not possible to observe full crystallization or melting in the time of the simulation.

\begin{figure}\begin{center}
  \includegraphics[width=0.9\columnwidth]{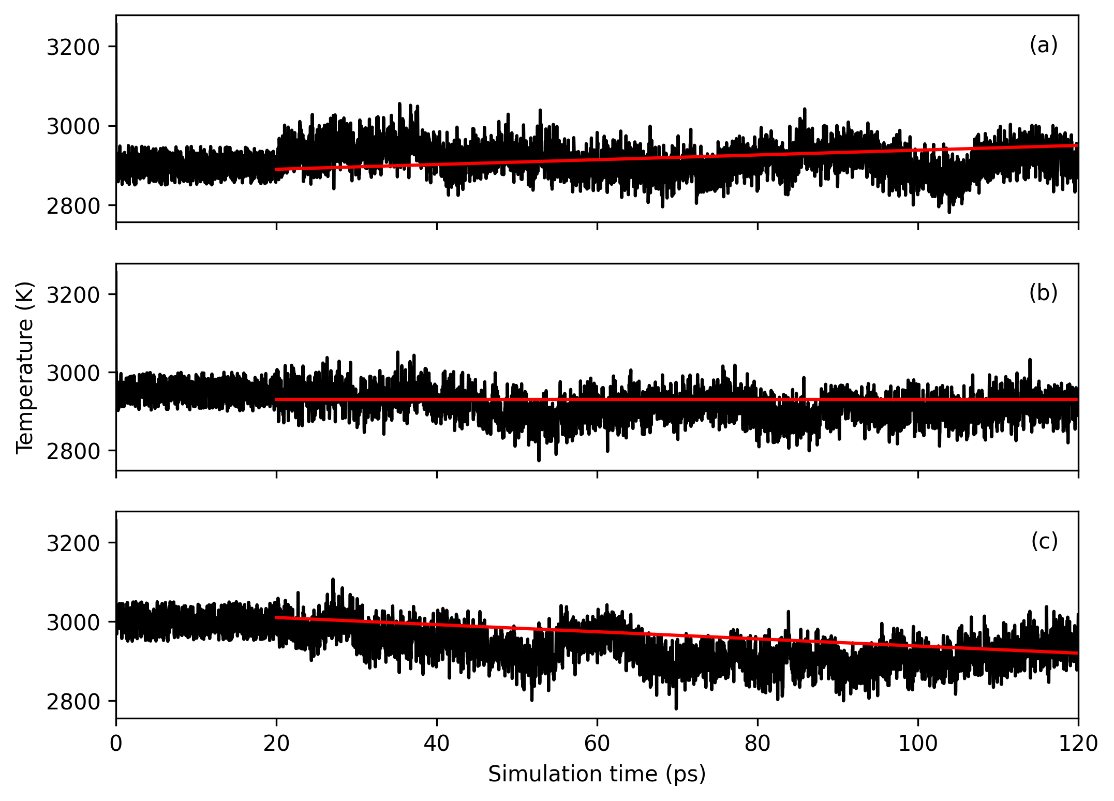}
  \caption{Temperature as a function of simulation time in a 6$\times$6$\times$12 supercell of \ce{CaO}, half in the solid phase and half in the liquid phase, at different initial temperatures: (a) 2900 K, (b) 2950 K, (c) 3000 K. The first 20 ps of the simulation are run in an NPT ensemble, while the subsequent 100 ps of the simulation are run in an NPH. The red lines represent an exaggeration of the linear regression of the curves, as a guide for the eye to see the global increase or decrease in temperature of the system. The actual linear regressions do reflect the trend of the curves, but the slope is not significant enough to be perceived.} 
  \label{fig:twophase}
\end{center}\end{figure}

At this point we run an NPT equilibration for 20~ps at different $T_{eq}$ close to the estimated $T_f$ followed by 100~ps in the isobaric-isenthalpic (NPH) ensemble and we monitor the temperature. There are three possible outcomes of the simulation: (i) if $T_{eq}<T_f$ the liquid portion of the simulation box tends to recrystallize (because it is an exothermic process) and the temperature increases throughout the simulation; (ii) if $T_{eq}>T_f$ the solid portion of the simulation box tends to melt (because it is an endothermic process) and the temperature of the system decreases throughout the simulation; (iii) obviously, if $T_{eq}\simeq T_f$ the liquid and the solid will coexist and the temperature of the system will remain constant throughout the simulation. The melting temperature can be extracted from the time average of this last MD run. This is shown in Fig.~\ref{fig:twophase}. The plots show that by considering different initial temperatures of 2900~K, 2950~K  or 3000~K  the system behaves differently.

In the first plot, in fact, the slight increase in temperature after the equilibration time shows that
the system tends to recrystallize, meaning that T=2900~K is below the melting point. On the other hand, the last panel shows an opposite behavior, since the global temperature between 20 and 120~ps of simulation time slightly decreases, meaning that T=3000~K is above the melting point. The small positive and negative slopes respectively of the trend line of these simulations show that both the initial temperatures of 2900~K and 3000~K are already quite close to the equilibrium temperature, so the melting temperature is to be found between these two values. In fact, the middle panel, which refers to a simulation with an initial external temperature of 2950~K, shows that the system is at equilibrium since the global temperature throughout the NPH simulation oscillates around the starting temperature.

As a result of these observations, we infer that the melting point of \ce{CaO} by the two-phase solid-liquid coexistence technique is $2940\pm65$~K, calculated as the average temperature of the NPH
simulation shown in Figure \ref{fig:twophase}b. Although a little bit lower than laser heating experimental measurements~\cite{Manara2014,Bgasheva2021}, the value of $T_f$ obtained is consistent with the value of $3066\pm12$~K which was obtained in this work by employing the void-nucleated melting technique.

\begin{figure}\begin{center}
  \includegraphics[width=0.9\columnwidth]{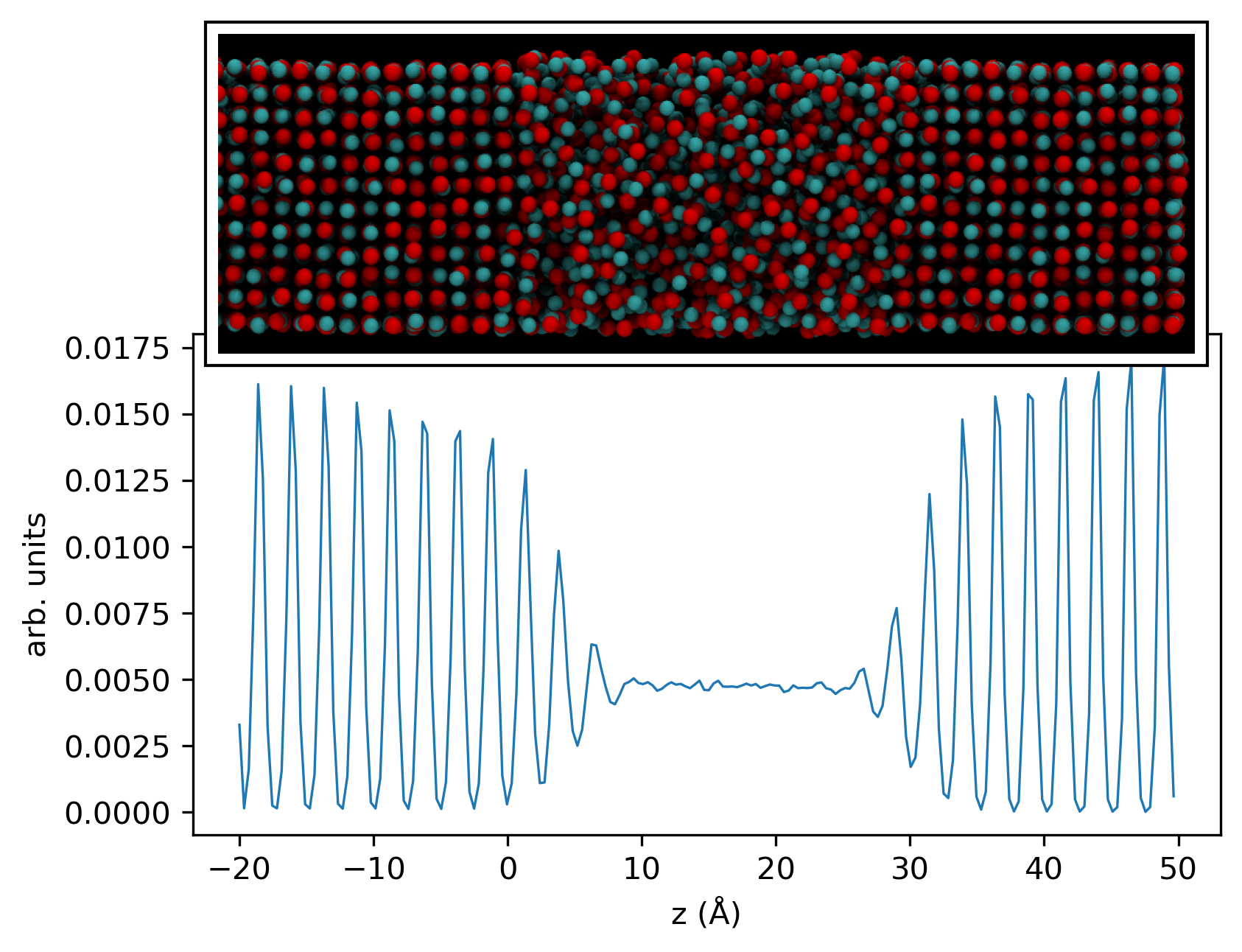}
  \caption{Snapshot of the simulation box with coexisting solid and liquid phases, visualized
  with VMD~\cite{Humphrey1996}. The 6$\times$6$\times$12 supercell is composed in part of the liquid phase (on the middle of the panel) and in part of the crystal phase (on the left-hand and the right-hand side of the panel). The red spheres represent oxygen atoms while the light blue ones represent calcium atoms.}
  \label{fig:snapshot}
\end{center}\end{figure}

Fig.~\ref{fig:snapshot} shows a snapshot of the simulation box at the ambient-pressure melting point. In the same Figure we also report the density profile averaged over the last 20~ps. Contrary to what is reported in Ref.~\cite{Wang2023}, the density profile of the liquid phase does not show any oscillation. Moreover, the solid-liquid interface extends over 4-5 atomic planes and the density profile of the bulk is recovered with 10--15~Å from the liquid. This is in agreement with previous MD simulations of the ionic compound NaCl solid-liquid and liquid-vapor interface~\cite{Zykova2005a,Zykova2005b}.

\subsection{Enthalpy of fusion}
The enthalpy of fusion ($\Delta H_f$) is the heat required for a crystal to change its state from solid to liquid, at constant pressure. The enthalpy of fusion for \ce{CaO} is quite difficult to constrain experimentally due to the very high temperatures required to run the experiments as well as to several technical issues related to the conditions of the sample and the experimental setup, as briefly outlined in Sec.~\ref{sec:introduction}. This is the reason why, as far as we know, there are no direct measurements of $\Delta H_f$ currently available for \ce{CaO}. To estimate $\Delta H_f$ we calculate the caloric curve at ambient pressure for the crystal and liquid branches by MD. \RED{Here we used both classical and ab initio MD}.

\RED{With CMD} we run several simulations at increasing temperature, starting from the crystal phase at T=300~K and gradually heating and equilibrating the system until we observe homogeneous melting. Then, we start again from the obtained liquid phase, and gradually lower the temperature from T=4000~K, cooling down stepwise until we observe a partial recrystallization of the liquid. Every point of the curve is the result of a NPT simulation employing a 6$\times$6$\times$6 supercell of 1728 atoms. Every simulation is 300,000 steps long with a timestep of 0.5~fs, which results in a total simulation time of 150~ps composed by 50~ps of equilibration and 100~ps of production. The solid phase is heated up stepwise with a temperature step of 50~K up to 4000~K, and then the resulting liquid is cooled down to ambient pressure at the same rate. The enthalpy is calculated as the average enthalpy value during the production stage for each simulation. 

\RED{Since AIMD simulations are rather expensive, we simulate a 3$\times$3$\times$3 cubic supercell of 216 atoms for 10,000 steps of 1~fs each. We perform dynamics in the canonical ensemble (NVT), employing the velocity-rescaling thermostat. Newton's equations of motion are integrated by the means of the Verlet algorithm. We perform AIMD at different temperatures, from ambient temperature up to approximately the expected melting temperature, namely at 300~K, 500~K, 1000~K, 1500~K, 2000~K, 2500~K, 3000~K and 3200~K for the crystal phase in the B1 structure and 3000~K, 3500~K and 4000~K for the liquid phase. At each temperature, the density of the system is fixed according to the average density of the solid/liquid obtained by classical MD.}

\begin{figure}\begin{center}
  \includegraphics[width=0.9\columnwidth]{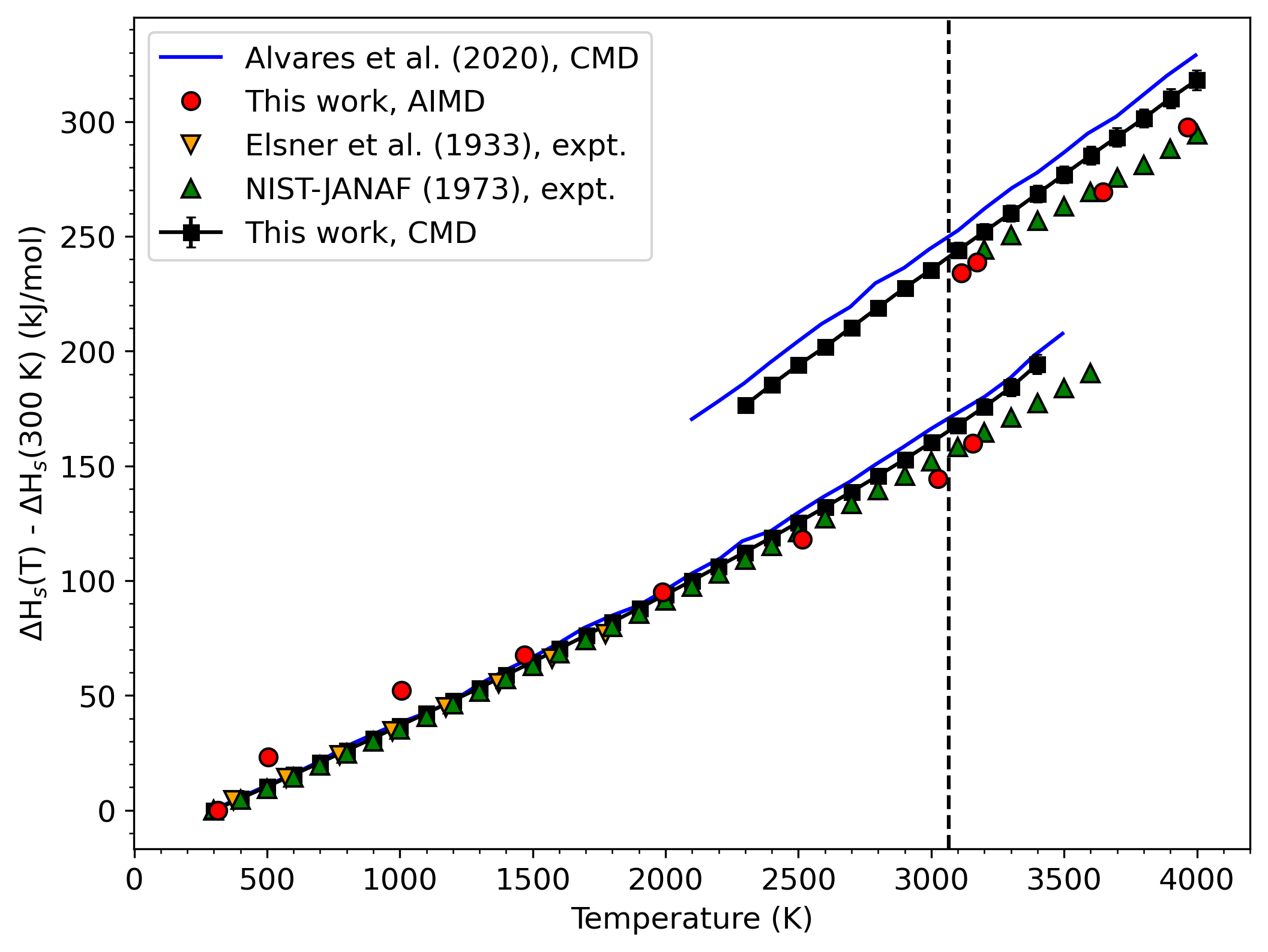}
  \caption{Enthalpy change (relative to the standard-state enthalpy of the solid, i.e. $H^0_{S,300}$) as a function of temperature. The lower branch refers to the crystal phase, while the upper branch refers to the liquid phase. The vertical dotted line corresponds to the melting temperature of $T_f=3066\pm12$~K, as obtained by the void-nucleated melting technique. Experimental and tabulated values are shown for comparison, along with the results of another MD simulation.}
  \label{fig:caloric}
\end{center}\end{figure}

The results are shown in Fig.~\ref{fig:caloric}. Solid and liquid enthalpy branches are plotted with respect to the reference temperature $T_0 = 300$~K and are truncated when the crystal phase reaches  the homogeneous melting (at 3450~K) and the liquid phase displays crystallization (at 2150~K) respectively in CMD.

\RED{Using CMD} the calculated enthalpy values of the solid branch are in excellent agreement with the experimental data obtained by drop-solution calorimetry below 2000~K (see Ref.~\cite{vonGronow1933} and references therein; see also Ref.~\cite{Chase1998}), while both branches are consistent with the existing data from molecular dynamics calculations~\cite{Alvares2020}. Our AIMD-PBEsol calculations for the solid phase fit well with the enthalpy data tabulated by NIST-JANAF~\cite{Chase1998} up to 2000~K, then slightly depart from the assessed trend at higher temperatures. At 3000~K, in fact, both AIMD simulations on the crystal phase and the liquid phase underestimate classical MD results by a maximum of 15\%. On the other hand, classical MD results obtained in this work are closer to the NIST-JANAF values for both the solid and the liquid phases with respect to Alvares et al.~\cite{Alvares2020}. \RED{AIMD calculations overestimate the enthalpy at low T and tend to slightly underestimate it close to the melting point. Given the higher cost of AIMD simulation in terms of both computational power and time, we performed AIMD simulations on a much smaller supercell of just 216 atoms, which can be the cause of the above-mentioned discrepancies.}

The enthalpy of fusion ($\Delta H_f$) is calculated as the enthalpy difference between the two curves at the melting temperature $T_f$. We take as reference the melting temperature we obtained employing the void-nucleated melting technique (i.e. $T_f=3066\pm12$~K), which yields a calculated enthalpy of fusion of 80.37~kJ/mol \RED{using CMD}. This value is in excellent agreement with the value retrieved from the NIST-JANAF Tables~\cite{Chase1998} ($\Delta H_f=79.5$~kJ/mol) and also perfectly fits the enthalpy of fusion obtained by Alvares et al.~\cite{Alvares2020} employing CMD ($\Delta H_f=80.89$~kJ/mol) (see Tab.~\ref{tab:enthalpy}). Further evidence supporting the robustness of the predicted $\Delta H_f$ values of \ce{CaO} by MD is the comparison with previous thermodynamic assessments, which provide enthalpies of fusion roughly between 80 and 76~ kJ/mol~\cite{Deffrennes2020,Belmonte2017a}, with a typical uncertainty of 10--20~kJ/mol for this kind of compounds~\cite{Ottonello2013,Belmonte2013,Belmonte2017b}. With AIMD we computed the solution enthalpy, i.e. the difference between the liquid and the solid at temperature other than the melting temperature. The results are 89.66~kJ/mol at 3000~K and 78.90~kJ/mol at 3200~K, which are in the range bracketed by previous calculations and experimental data.

\begin{table}\begin{center}
\begin{tabular}{ccc}
\hline\hline
    $\Delta H_f$ (kJ/mol) & Method & Reference \\
    \hline
    80.37 & CMD &This work \\
    89.66 & AIMD@3000~K& This work \\
    78.90 & AIMD@3200~K&This work \\
    80.89 & CMD & Alvares et al.~\cite{Alvares2020}\\
    80 $\pm$ 16 & therm. assesment & Deffrennes et al.~\cite{Deffrennes2020} \\
    79.5  & estimated from MgO & NIST-JANAF Tables~\cite{Chase1998} \\
\hline\hline
\end{tabular}
\caption{Values of the enthalpy of fusion of \ce{CaO} in kJ/mol as obtained from different sources.}
\label{tab:enthalpy}
\end{center}\end{table}

\begin{figure}\begin{center}
  \includegraphics[width=0.9\columnwidth]{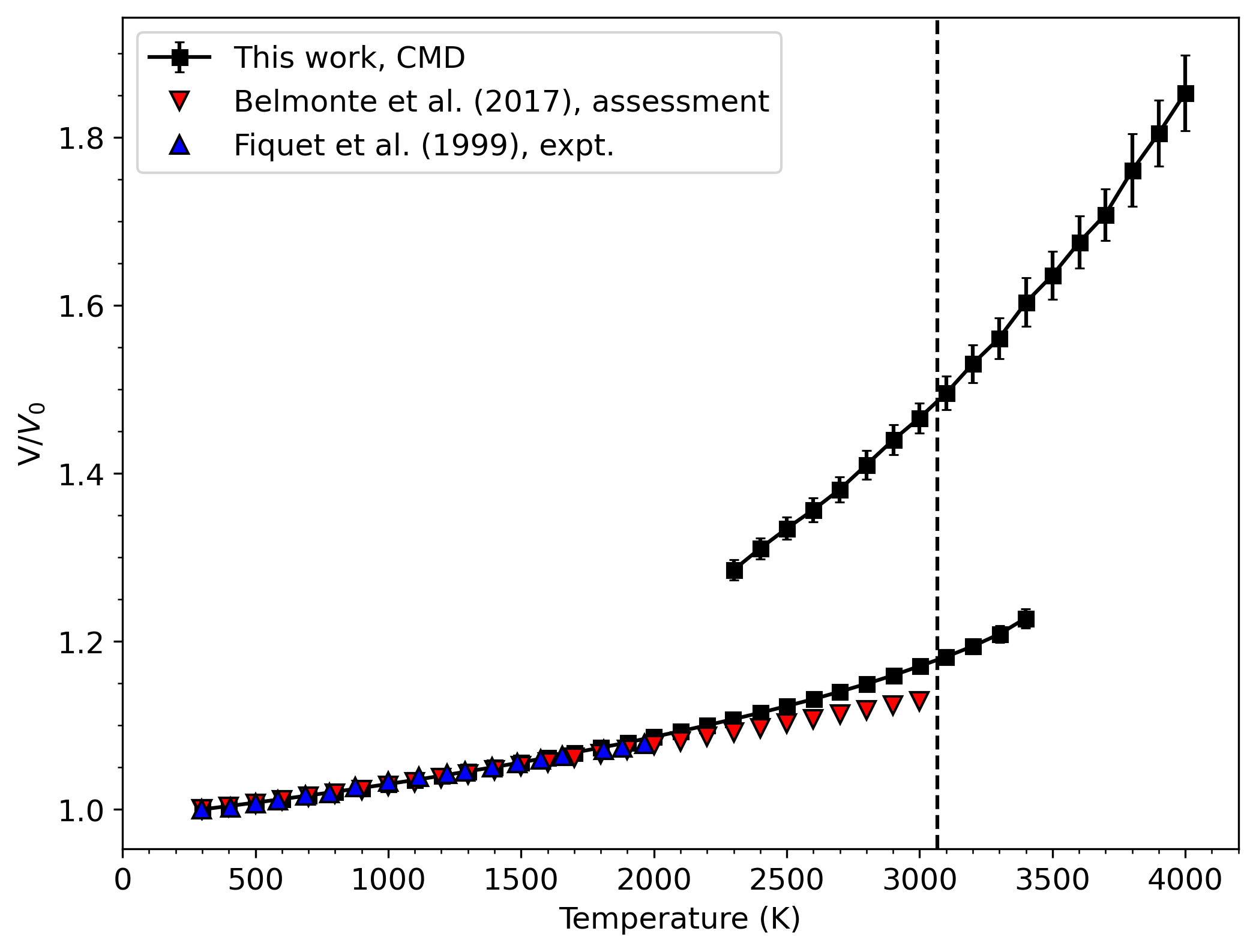}
  \caption{Relative volume thermal expansion as a function of temperature. The lower branch refers to the crystal phase, while the upper branch refers to the liquid phase. The vertical dotted line corresponds to the melting temperature of $T_f=3066\pm12$~K, as obtained by the void-nucleated melting technique. Experimental~\cite{Fiquet1999} and other tabulated values~\cite{Belmonte2017a} are shown for comparison.}
  \label{fig:volume}
\end{center}\end{figure}

\RED{In addition to the caloric curve, we report in Fig.~\ref{fig:volume} the relative volume thermal expansion as a function of temperature, at ambient pressure, using CMD. Our results are basically indistinguishable form those of Alvares et al.~\cite{Alvares2020}. For the sake of comparison, we added experimental data~\cite{Fiquet1999} and results from previous DFT calculations~\cite{Belmonte2017a}. Although the BMH potential slightly underestimates the lattice parameter at low T, the lattice thermal expansion is very well reproduced. The larger thermal expansion above 2500~K is due to the appearance of high-order anhamornic terms in MD, which are not fully accounted for by the ab initio Quasi Harmonic Approximation. The volume jump between the solid and the liquid is about 25\% which is typical of ionic systems~\cite{Zykova2005a}.}

\subsection{High-pressure melting curve }
The results reported so far demonstrate that the two-phase coexistence method provides a $T_f$ that is consistent with the void-nucleated technique, using an interatomic potential validated on first principles calculations~\cite{Alvares2020}. Moreover, the caloric curve also agrees very well with thermodynamic tabulations and CMD results. In this section we apply the two-phase coexistence method to calculate the melting curve of \ce{CaO} as a function of pressure. This is because the void-nucleated technique can be used up to relatively low pressures, until the crystal with the void mechanically collapses.

The first attempt to calculate the melting curve of \ce{CaO} was made by Sun et al.~\cite{Sun2010}. They employed classical MD up to 60~GPa of pressure to obtain the thermal instability temperature ($T_s$) of \ce{CaO}, which is known to overestimate the actual melting temperature ($T_f$) of about 20\% for ideal crystals~\cite{Jin2001}. Since they found $T_s$ at ambient pressure which was about 30\% higher than the value of $T_f=3200\pm50$~K which they assumed from the NIST-JANAF Thermochemical Tables the high-pressure melting curve of \ce{CaO} was estimated in that work by applying a 30\% correction on the computed thermal instability temperature, for every value of the pressure.

In this work, we employ \RED{CMD and} the two-phase solid-liquid coexistence technique to obtain the high-pressure melting curve of \ce{CaO} up to 30~GPa, limit above which the lack of a short-range repulsive barrier in the interatomic potentials lead to the failure of the MD integration. We run the simulations employing the same empirical potential used for previous MD simulations, using a 6$\times$6$\times$12 supercell of 3456 atoms. First, we run 20~ps of an isobaric-isothermal (NPT) simulation to equilibrate the system at the desired conditions of temperature and pressure. We follow the procedure described earlier. When the temperature is close to $T_f$ and the solid-liquid interface doesn't move, we perform 100~ps of NVH simulation so as to obtain $T_f$ as the time average of the temperature. In order to get the thermal instability temperature curve ($T_s$) instead, we perform NPT simulations on a 6$\times$6$\times$6 supercell of 1728 atoms without defects, heating the perfect crystal at a temperature rate of $10^{12}$~K/s for a total simulation time of 250~ps. Since the supercell has no defects, the melting temperature observed in that case is affected by overheating, and thus fairly represents $T_s$. We run both type of simulations increasing the external pressure stepwise from ambient pressure to 30~GPa, with a pressure step of 5~GPa.

\begin{figure}\begin{center}
  \includegraphics[width=0.9\columnwidth]{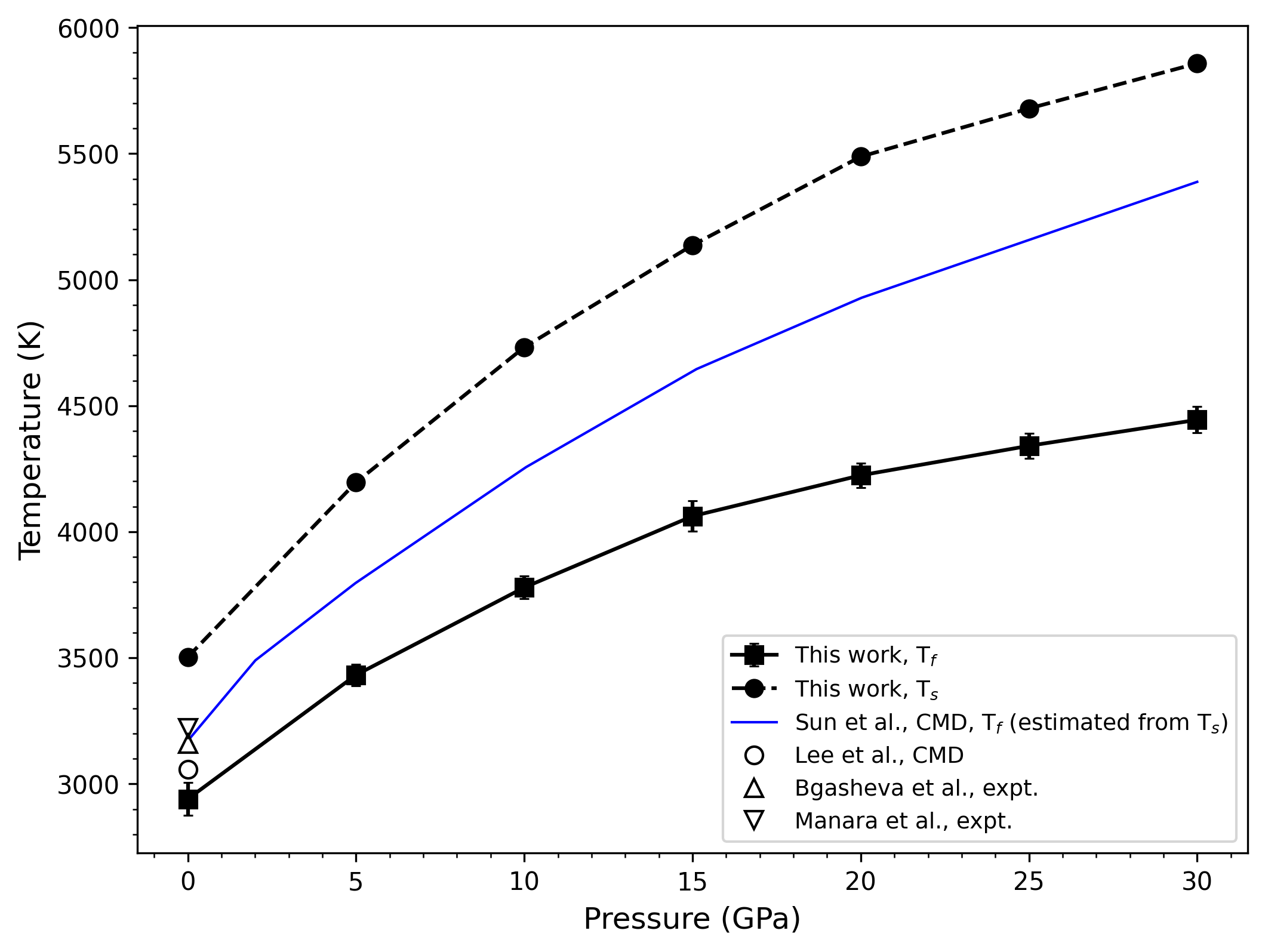}
  \caption{High-pressure melting curve of \ce{CaO} ($T_f$, black squares) and thermal instability temperature curve ($T_s$, black dots) as calculated in this work. The melting curve estimated by Sun et al.~\cite{Sun2010} is shown for comparison (in blue). Empty symbols represent different values of $T_f$ obtained at ambient pressure by laser heating experiments or MD simulations.}
  \label{fig:melting_curce}
\end{center}\end{figure}

Fig.~\ref{fig:melting_curce} shows the results of MD simulations performed in this work together with the available results in the literature. The high-pressure melting curve of \ce{CaO}, which is rather steep at lower pressures, rapidly flattens with increasing pressure. It is interesting to note, though, that the two curves do not have the same slope since $T_s$ increases more rapidly with pressure than $T_f$. In fact, the overheating ratio $\eta=\frac{T_s}{T_f}-1$ increases as the pressure increases, going from about 16\% at ambient pressure to more than 24\% at P=30~GPa. Therefore the assumption made by Sun et al.~\cite{Sun2010} to obtain the melting curve of \ce{CaO} at high pressure conditions by means of a constant scaling factor is not justified from a thermodynamic point of view and their estimated values of $T_f$ from $T_s$ are affected by an uncertainty of at least 10--15\%. 

\begin{table*}\begin{center}
\begin{tabular}{cccc}
\hline\hline
    Pressure (GPa) & $T_f$ (K) & $T_s$ (K) & Overheating ratio (\%) \\
    \hline
    0  & 2940$\pm$65 & 3503 & 19.1 \\
    5  & 3432$\pm$43 & 4196 & 22.3 \\
    10 & 3779$\pm$45 & 4732 & 25.2 \\
    15 & 4062$\pm$60 & 5137 & 26.5 \\
    20 & 4224$\pm$48 & 5488 & 29.9 \\
    25 & 4341$\pm$49 & 5679 & 30.8 \\
    30 & 4444$\pm$52 & 5858 & 31.8 \\
\hline\hline
\end{tabular}
\caption{Calculated values for $T_f$ and $T_s$ of \ce{CaO} 0 to 30~GPa and the corresponding overheating ratio (in \%)}
\label{tab:melting}
\end{center}\end{table*}

In Tab.~\ref{tab:melting} we report all the calculated values for $T_f$ and $T_s$ at different pressure conditions, along with the corresponding overheating ratio. The two quantities that we obtained from classical MD can be fitted with the following empirical equations:
\begin{eqnarray}
  T_f &=& 235.241\, P^{0.558} + 2926.195 \nonumber\\
  T_s &=& 300.369\, P^{0.618} + 3480.601 \label{eq:formulas},
\end{eqnarray}
\RED{where the temperature ($T_f$ and $T_s$) is given in K and the pressure ($P$) in GPa.}

The rationale behind the divergence between the melting and the thermal instability temperatures with increasing pressure is discussed in depth by G\'omez et al.~\cite{Gomez2005} who employed classical MD to investigate Lennard-Jones crystals and found the very same phenomenon we observe for \ce{CaO}. This can be explained by empirical laws described in which have the same form of the Equations 2 and 3 above. While the exponents of the pressure in these laws are similar, the key role is played by the two multiplicative constants: the larger prefactor of $T_s$ in Eq.~\ref{eq:formulas} can be explained by the lack of nucleation sites in the perfect crystal. In absence of defects or interfaces, the appearance of clusters of defects that break the crystal structure (i.e. liquid-like regions) require a high thermal activation barrier. This activation barrier increases as pressure increases, due to the higher binding energy between atoms. In the presence of a solid-liquid interface, the process involved is different since the crystalline phase is put in contact with its own liquid phase from the very beginning of the simulation. In this case, the solid-liquid interface does not experience a dramatic increase in the energy barrier for moving, since the external pressure leads to an increased binding energy mostly in the liquid, which is more compressible than the solid. Therefore, assuming a constant overheating ratio of $\sim$30\% between $T_s$ and $T_f$ is not a valid assumption, especially at low pressures.

%%%%%%%%%%%%%%%%%%%%%%%%%%%%%%%%%%%%%%%%%%%%%%%%%%%%%%%%%%%%%%%%%%%%%%%%%%%%%%%%%%%%%%%%%%%%%%%%%%
\section{Conclusions}
%%%%%%%%%%%%%%%%%%%%%%%%%%%%%%%%%%%%%%%%%%%%%%%%%%%%%%%%%%%%%%%%%%%%%%%%%%%%%%%%%%%%%%%%%%%%%%%%%%
The melting point of \ce{CaO} (lime) is currently known from experiments with a large uncertainty ($\sim$400~K) depending on the employed technique. Despite the large number of studies, this discrepancy is not yet resolved. Moreover, the melting curve of CaO as a function of pressure has not been determined, to date.

We performed classical MD simulations with an ab-initio derived interatomic potential~\cite{Alvares2020}. We compared two methods for calculating the melting temperature at ambient pressure. Our results for $T_f$ are 3066$\pm$12~K with the void-nucleated technique and 2940$\pm$65~K with the two-phases solid-liquid coexistence technique. These values are consistent with the most recent experimental data. Next, using classical MD, we calculated the enthalpy of fusion $\Delta H_f$ to be 80.37~kJ/mol, which is consistent with assessed data in the literature. Finally, using the two-phase coexistence method, we calculated the melting curve of \ce{CaO} up to 30~GPa without making any assumption on the Clapeyron slope, and compared it to the thermal instability limit $T_s$. Our results show that $T_s$ increases more steeply than $T_f$ with increasing pressure and assuming a constant $T_s/T_f$ ratio (as it is often done in literature~\cite{Sun2010}) is not a valid assumption. Thus our results provide an extremely valuable insight for the thermodynamic assessment of melting phase relations in \ce{CaO}-bearing multi-component oxide systems of geochemical relevance up to high pressure conditions.

\begin{acknowledgement}
The Authors acknowledge the CINECA award under the ISCRA initiative (ISCRA C Projects HP10CA2JUN and HP10C2UBUG) for the availability of HPC resources and support. DB also acknowledges the financial support from the Italian MIUR PRIN 2017 (project number: 2017KY5ZX8) and MUR PRIN 2020 (project number: 202037YPCZ).
\end{acknowledgement}

\bibliography{references}

%%%%%%%%%%%%%%%%%%%%%%%%%%%%%%%%%%%%%%%%%%%%%%%%%%%%%%%%%%%%%%%%%%%%%%%%%%%%%%%%%%%%%%%%%%%%%%%%%%
\end{document}